\documentclass[acmlarge]{acmart}
\makeatletter
\newcommand{\confshort}{\acmConference@shortname}
\newcommand{\conffull}{\acmConference@name}
\newcommand{\confdate}{\acmConference@date}
\newcommand{\confloc}{\acmConference@venue}
\AtBeginDocument{
  \fancypagestyle{firstpagestyle}{
    \fancyhead{}%
    \fancyfoot[C]{}%
  }
  \fancyhf{}
  \fancyhead[LO]{\@headfootfont\shorttitle}%
  \fancyhead[RE]{\@headfootfont\@shortauthors}%
  \fancyhead[LE]{\@headfootfont\footnotesize \confshort, \confdate, \confloc}%
  \fancyhead[RO]{\@headfootfont\footnotesize \confshort, \confdate, \confloc}%
  \fancyfoot[C]{}%
}
\makeatother
\acmBooktitle{\conffull\@ (\confshort), \confdate, \confloc}

\AtBeginDocument{%
  }

\copyrightyear{2026}
\acmYear{2026}
\acmDOI{10.1145/3805689.3806506}
\setcopyright{cc}
\setcctype{by-nc-nd}
\acmConference[FAccT '26]{The 2026 ACM Conference on Fairness, Accountability, and Transparency}{June 25--28, 2026}{Montreal, QC, Canada}
\acmBooktitle{The 2026 ACM Conference on Fairness, Accountability, and Transparency (FAccT '26), June 25--28, 2026, Montreal, QC, Canada}
\acmISBN{979-8-4007-2596-8/2026/06}

\usepackage[capitalise,noabbrev]{cleveref}
\usepackage{microtype}
\usepackage{tabularx}
\usepackage{booktabs}
\usepackage{enumitem}

\begin{document}

\title{AI, Digital Platforms, and the New Systemic Risk}

\author{Philipp Hacker}
\email{hacker@europa-uni.de}
\affiliation{%
  \institution{European University Viadrina}
  \city{Berlin}
  \country{Germany}}

\author{Lilian Edwards}
\email{Lilian.Edwards@newcastle.ac.uk}
\affiliation{%
  \institution{University of Glasgow}
  \city{Glasgow}
  \country{UK}}

\author{Atoosa Kasirzadeh}
\email{atoosa.kasirzadeh@gmail.com}
\affiliation{%
  \institution{Carnegie Mellon University}
  \city{Pittsburgh}
  \country{USA}}

\begin{abstract}
Artificial intelligence (AI) is becoming increasingly embedded in digital, social, and institutional infrastructures, often merging with platforms to form hybrid structures. As a result, systemic risk has emerged as a critical but undertheorized challenge. In this paper, we develop a rigorous framework for understanding systemic risk in AI, platform, and hybrid system governance. We draw on insights from finance, complex system theory, and cybersecurity -- domains where systemic risk has already shaped regulatory responses. We argue that recent legislation, in particular the EU's AI Act and Digital Services Act (DSA), invokes systemic risk but relies on narrow or ambiguous characterizations of this notion. At times, the legislation reduces the source of risk to specific capabilities in frontier AI models; at other times, it attributes the risk to harms occurring in economic market settings. The DSA, we posit, actually does a better job at identifying systemic risk than the more recent AI Act. Our framework identifies systemic risks overlooked by the DSA and AI Act --- including multi-agent interactions, discrimination at scale, and large-scale hallucinations --- which can destabilize institutions despite falling outside current definitions. More generally, we present a framework for systemic risk, categorized into four levels: single-model, multi-model, model-platform, and model-institution integration. We then test the characterizations of systemic risk in DSA, the AI Act, and our own framework by applying them to three key yet contentious examples -- large-scale discrimination, hallucinations, and environmental effects -- to illustrate their respective strengths and limitations. Against this background, this paper proposes reforms that broaden systemic risk assessments, strengthen coordination between regulatory regimes, and explicitly incorporate collective harms. As localized AI and platform failures may increasingly escalate into structural disruptions, we provide a conceptual foundation and a policy-relevant diagnostic toolkit for governing the interplay of AI and platforms in complex, interconnected societies.
\end{abstract}

\ccsdesc[500]{Social and professional topics~Computing / technology policy}
\ccsdesc[500]{Social and professional topics~Governmental regulations}
\ccsdesc[300]{Computing methodologies~Artificial intelligence}

\keywords{systemic risk, AI governance, Digital Services Act, AI Act, platform regulation, fundamental rights}

\maketitle

\section{Introduction}

Throughout history, societies have grappled with technologies that promise progress yet can destabilize the systems into which they are introduced. This tension made the 20th century a period of sharper differentiation between individual risk, defined as specific harms limited to a single person or a few individuals, and systemic risk, characterized by a structural nature that warrants distinct analysis and, at times, regulation. The 2008 financial crisis crystallized decades of research in financial economics into a global event that destabilized economies and drew urgent attention to the concept of systemic risk. Today, systemic risks to financial infrastructure have not disappeared. Yet the growing complexity of digital infrastructure, and its entanglement with core political and social processes, adds new dimensions of systemic risk.

Historically, systemic risk has been well defined in finance: cascading failures that destabilize the financial system as a whole (see Section~\ref{sec:historical}). AI and large platforms broaden this risk landscape, creating common exposures and points of failure that traditional formulations do not fully capture. This shift is now reflected in digital and AI risk governance, which increasingly treats systemic risks as a critical concern. Recent legislative efforts, notably the European AI Act and the Digital Services Act (DSA), reflect this growing concern.

The DSA identifies four major categories of systemic risk that providers of very large digital platforms and search engines must mitigate: the dissemination of illegal content, negative effects on fundamental rights, risks to civic discourse and electoral processes, and harms related to minors and public well-being~\cite{marsh2024systemic}. The EU AI Act establishes obligations for providers of general-purpose AI (GPAI) models regarding systemic risks. Unlike the DSA’s enumerative approach, the AI Act provides a conceptual definition focusing on high-impact capabilities, market-wide effects, and propagation across the value chain (Art. 3(65)).

In practice, the AI Act attempts to designate AI models as possessing systemic risks based primarily on training compute thresholds, presuming ``high-impact capabilities'' when training involves more than $10^{25}$ floating-point operations (FLOPs). By this measure, many large language models (LLMs) on the market would be classified as systemically risky. However, as \citet{bertuzzi2025inside}, \citet{hacker2026regulation}, and \citet{somala2025three} note, rapid advancements in reinforcement learning and compute efficiency have challenged this designation method even before proper implementation. If systemic risk is defined solely by compute thresholds, the Act may overlook the contribution of smaller AI systems deeply integrated into critical infrastructure, while misclassifying computationally intensive models with limited societal integration as systemic risks. Consequently, the Act's compute threshold is already under scrutiny for revision.

More specifically, Art. 3(64) of the AI Act defines high-impact capabilities as ``capabilities that match or exceed the capabilities recorded in the \emph{most advanced} general-purpose AI models [our emphasis].'' Under this logic, systemic risks must be \emph{specific} to these frontier models. However, it is not clear why risks such as large-scale manipulation, discrimination, or hallucinations should only be considered "systemic" when they stem from the most advanced systems; such harms can arise across a wide range of AI models and still cause widespread disruption. This ambiguity highlights the need for a structured framework that transcends simple size-based compute thresholds. While Annex XIII of the AI Act provides a broader set of designation parameters, these remain deeply connected to a definition of systemic risk that focuses exclusively on “the most advanced” models, as we shall discuss later in the paper.

Despite these regulatory mentions, the boundaries of systemic risk in AI-driven systems and platforms remain underdeveloped. Systemic risk is often regarded as a ``you-know-it-when-you-see-it'' concept~\cite{benoit2017risks,marsh2024systemic}. This stance, we argue, is problematic and can render the notion so loose and useless as to provide little practical guidance: to clarify systemic risk in AI-driven systems, platforms, and hybrid structures that combine both requires one to identify, measure, and monitor the complex propagation mechanisms by which localized failures could cascade across interconnected systems. If systemic risk is defined exclusively by the advanced capabilities of single AI models, the definition contradicts decades of research on systemic risks in domains like finance, where the focus lies on interaction effects rather than the properties of individual components.

In this paper, we develop an analytic framework for understanding systemic risk in digital governance by identifying conditions under which AI or platform risks transition from localized failures to systemic risks. To achieve this, we proceed in four steps. First, we draw on lessons from the historical study of systemic risk, financial risk regulation, and complex system sciences to identify minimal conditions common to core definitions of systemic risk in the literature (Section~\ref{sec:historical}). Second, and based on our initial historical findings, we formulate a conceptually rich list of four assessment criteria for specifying AI- and platform-driven systemic risks (Section~\ref{sec:conceptual}): (i) the scale and scope of expected damage; (ii) the emergence of collective harms that exceed the sum of individual impacts; (iii) the potential irreversibility of certain harms, and (iv) complexity---for example interconnectedness--- which enables cascading and non-contained effects. Together, they capture the dimension (i), the nature (ii and iii) and the unpredictability (iv) of harm from the manifestation of systemic risk. Third, we identify four key levels at which systemic risks from AI- and platform-driven systems could manifest: (1) single-model systemic risks: failures from a single entity (AI model or platform), when widely adopted; (2) multi-model systemic risks: systemic failures when multiple AI models fail in synchronized ways; (3) model-platform integration systemic risks: AI systems embedded into large-scale platforms that create feedback loops; (4) model-institution integration systemic risks: AI systems embedded into governance, law enforcement, and finance that create structural risks.

Finally, we apply our investigation to AI and digital risk governance (Sections~\ref{sec:dsa}--\ref{sec:examples}). In particular, we discuss the merits of addressing systemic risks in the DSA and the EU AI Act; we argue that the interpretation of systemic risk in the AI Act is too narrow, that additional governance and monitoring mechanisms are required to contain potential systemic failures, and that a more holistic understanding of systemic risk is necessary to tackle hybrid structures combining AI and platforms.

\section{Related Work}

Much prior work has criticized the concept of systemic risk as a regulatory target in EU digital regulation, including the Digital Services Act (DSA) and subsequently the AI Act. In particular, the DSA is criticized for \emph{lacking a core definition} of systemic risks and instead providing an open-ended list, which provides flexibility but leaves interpretation open to political pressure and platform provider negotiation~\cite{griffin2025politics,kaesling2025sustainability,loi2025regulating}. Griffin~\cite{griffin2025politics} emphasises that
translating broad, value-laden concepts into concrete risk mitigation measures will necessarily
involve \emph{normative choices and power struggles} among regulators, platforms, civil society,
and experts. Because systemic risk is framed as stemming from the ``design or functioning of
the service and its related systems,'' a central interpretive question is: \emph{which system(s)} are we
assessing – the platform as a whole, specific recommender systems, ad delivery stacks, or broader digital ecosystems? Kaesling and Wolf~\cite{kaesling2025sustainability} use sustainability risks to show how hard it is to decide when a risk is truly ``systemic'' in the DSA sense versus a more local or isolated harm. Critics also note that because the DSA concept of systemic risk overlaps with other regimes including financial rules, media and electoral rules and now the AI Act, fragmented, \emph{conflicting and/or duplicative risk assessments} may become a key practical problem possibly stifling innovation~\cite{karathanasis2025regulatory,torraco2025risk}.

Prior scholarship on ``systemic risk'' as a concept in the AI Act also makes multiple
criticisms including: \emph{conceptual overbreadth and vagueness} - the Act blends different
conceptions of risk (technical risk, social risk, uncertainty) into a single ``systemic risk''
construct, which makes it difficult to operationalise and measure~\cite{griffin2025politics}; \emph{model vs system level}: systemic risk is defined and operationalised at the GPAI \emph{model} level, while many harms materialise at the \emph{system} or application level [2]; \emph{designation criteria and thresholds}: reliance
on compute-based thresholds and undefined ``high-impact capability'' benchmarks introduces
ambiguity about which models qualify and could lead to uneven, politically influenced
designations~\cite{karathanasis2025regulatory}; ``\emph{risk theatre}'' or ``\emph{appearance'' of control}: systemic risk obligations may
give an appearance of strong safety governance while leaving the actual determination and
scope of systemic risks uncertain~\cite{care2024central,karathanasis2025regulatory}.

Several authors also compare the DSA and AIA formulations to prior financial regulation in
terms of vagueness, lack of precise technical metrics, and ultimately questions as to who controls the notion and for what
aim. Compared to
finance, both Keller et al.~\cite{keller2023european} and Kutterer~\cite{kutterer2024regulating} suggest that AI Act risk concepts remain
predominantly micro or meso-level, not macro, and that the new ``systemic risk'' label for GPAI is more about precaution under uncertainty, than about identified system stability threats
in a clearly defined ``AI system.'' Loi, Fabbri and Ferrario~\cite{loi2025regulating} introduce the notion of a \emph{convergence problem}:
different disciplines, methods and data sources (e.g. computer science, social science, risk
engineering, legal doctrine) can legitimately identify and model systemic risks in
incompatible ways, and neither regulators nor platforms have a principled way to force
convergence on a single narrative. Palumbo~\cite{palumbo2025systemic} raises \emph{constitutional concerns}, given the
vagueness of the systemic risk framework, about \emph{how much} political discretion is being
delegated and to whom. Compared with financial regulation, where macro‑prudential
systemic‑risk tools have been anchored in relatively explicit mandates and indicator sets,
Palumbo worries that DSA/AIA systemic‑risk management rests on thinner democratic
grounding. This draws on work by O’Halloran and Nowaczyk which claims to show how
systemic risk can be made technically \emph{operationalizable} in finance, highlighting by contrast how underspecified ``systemic risk'' currently is in the DSA and AI Act~\cite{o2019artificial}.

Our own conceptual contribution builds on this related work, and extends it via a fourfold analysis: we offer a combination of (i) a historically grounded four-criteria framework for systemic risk, (ii) a four-level taxonomy of AI-specific risk pathways, (iii) a detailed comparative analysis of the DSA and AI Act systemic risk regimes, including the AI Act's frontier-model limitation, and (iv) a systematic application to three illustrative cases that tests both legal frameworks against our proposed conception.

\section{A Historical Perspective: Systemic Risk in Financial Markets}
\label{sec:historical}

The concept of systemic risk gained prominence following the Great Depression, but it was formally crystallized in financial literature during the 1980s and 90s, generally by reference to internal or external shocks which propagate through the market and affect a variety of financial or market institutions~\cite{allen2013systemic,brimmer1989distinguished,cline1984international,debandt2000systemic,galaz2021artificial,micova2023elements,summer2003banking}.


The concept of systemic risk entered the policy arena most notably with the Lamfalussy Report in 1990~\cite{bis1990lamfalussy}. 
It influenced subsequent regulatory frameworks for financial regulation, which in the 1990s addressed the mitigation of systemic risk that arose from interdependencies in the banking system. For example, the Settlement Finality Directive (98/26/EC) of 1998 aimed to reduce legal and systemic risks in payment and securities settlement systems.

Systemic risk became particularly embedded in EU financial regulation after the financial crisis of 2008.
Article 2(c) of the ESRB Regulation (EU) No 1092/2010 defines systemic risk as ``a risk of disruption in the financial system with the potential to have serious negative consequences for the internal market and the real economy.'' Systemic risk, in this understanding, is marked by the potential to threaten the stability of an entire system (in this case, the financial system), and the severity of harm for actors in that system in case of materialization. Importantly, systemic risk, therefore, has both a collective and an individual dimension. Its realization would have a severe negative impact on entities in the system, but would also destabilize the system as such.

These observations motivated \citet{kaufman2003what} to provide one of the most cited definitions: ``the risk or probability of breakdowns in an entire system, as opposed to breakdowns in individual parts or components.'' This definition emerged from studying the 1930s bank runs, where \citet{debandt2000systemic} documented how local bank failures triggered cascade effects across the entire financial system. The 2008 financial crisis led to a refinement by \citet{schwarcz2008systemic}, who emphasized that systemic risk represents ``the probability that cumulative losses will occur from an event that ignites a series of successive losses along a chain of institutions or markets comprising a system.''

Historically, scholars have thus conceptualized systemic risk primarily in terms of financial institutions, particularly banks. From this field, regulatory scholarship and practice can learn that sources of risk (shocks) can be external and internal; and effects are either idiosyncratic (largely limited to one entity) or systemic (affecting multiple actors)~\cite{debandt2000systemic,micova2023elements}. 

While financial systemic risk provides a clear, quantitative precedent for understanding how liquidity shocks and leverage contribute to market instability, the unique architecture of AI systems demands an even more expansive engagement with complex systems theory. Financial markets, despite their complexity, operate within relatively well-defined parameters of capital and regulation; conversely, AI ecosystems are characterized by higher degrees of opaque modularity and reflexive feedback loops, where emergent risks can bypass traditional equilibrium-based controls. As noted in the literature , systemic risk is not merely an aggregation of localized failures but an emergent property of high-order connectivity, where tight coupling allows minor perturbations to cascade into catastrophic transitions \cite{lucas2018systemic, renn2022systemic, renn2022systemic}. Understanding systemic risk through the lens of complex systems theory has necessitated new governance approaches, as traditional, reductionist models often fail to capture cascading failures. Consequently, frameworks like the IRGC guidelines now advocate for strategies that foster self-organization, iterative adaptation, and heightened preparedness for systemic shocks \cite{IRGC2018}. While these high-level guidelines provide a valuable starting point, we must first rigorously characterize the nature of these risks in the AI context before applying such guidelines to effectively identify the structural brittleness and potential for cascading contagion inherent to autonomous, large-scale deployments.

\section{Conceptual Dimensions of Systemic Risk}
\label{sec:conceptual}
From several definitions of systemic risk across complexity sciences, nuclear safety, and finance, collected in the Appendix, four main conceptual \textit{dimensions} emanate. 

\textbf{Scale and scope of harm.} Definitions emphasize the scale and scope of systemic risks. Unlike individual-level risks, systemic risks affect large portions of an entire system.

\textbf{Nature of harm I: Simple aggregation vs. collective harms.} Definitions represent a structural failure that at times transcends the cumulative total of individual harms, manifesting instead as a systemic collapse of social or institutional functions.

\textbf{Nature of harm II: Potentially irreversible.} Definitions emphasize the long-term and potentially irreversible nature of systemic risks.

\textbf{Unpredictable/not fully predictable harm: Complexity.} Definitions highlight the propagation and cascading nature of systemic risks. Complex interactions between components or systems create (partially) unpredictable outcomes that cannot be understood through simple linear models of cause and effect. 

Additionally, we posit that there are four \textit{levels} at which AI systemic risk can manifest:

\begin{enumerate}
    \item Single-model systemic risks -- failures from the pervasive adoption of a single, centralized AI model or platform across critical infrastructure, creating a monoculture effect. Because the same foundational entity underpins hundreds of downstream applications, a single point of failure can lead to widespread, synchronized disruption. For instance, if a leading language model produces systematically biased outputs, these errors are not contained; rather, they propagate across disparate sectors like hiring, lending, and content moderation, distorting decision-making at a scale that threatens the integrity of entire institutional processes \cite{hacker2025generative,bommasani2022picking}.
    \item Multi-model systemic risks -- correlated failures when multiple models fail in synchronized, or connected, ways. Such vulnerabilities materialize when various institutions deploy models trained on similar underlying datasets or architectures, leading to herding behavior that amplifies market volatility during periods of stress. Beyond financial applications, this systemic threat includes the potential for parallel or coordinated failures across a multitude of autonomous agents, or the emergence of cross-platform systemic fragility where multiple independent services share identical failure modes \cite{hammond2025multi,bisconti2025beyond,cemrimulti,hacker2026pragmatic}. Multiple platforms with similar failure modes could be another instance.
    \item Model-platform integration systemic risks -- AI embedded into large-scale platforms that generate feedback loops or spread harmful behavior. The recent case of Grok producing NCII at scale on X provides a nearly perfect example \cite{bbc_grok_iwf_2026}. More generally, these risks emerge when recommendation algorithms on social media platforms optimize for engagement metrics, which creates echo chambers that may erode public discourse, democratic deliberation, and respect for individual and collective rights \cite{cinelli2021echo}.
    \item Model-institution integration systemic risks -- AI systems embedded into high-stakes governance, finance, or law enforcement processes that create structural vulnerabilities. This occurs, e.g., when predictive policing systems encode historical biases into resource allocation decisions, which entrenches discriminatory patterns in criminal justice outcomes \cite{alikhademi2022review}.
\end{enumerate}

We now turn to evaluating two key EU legal acts designed to mitigate systemic risk: the DSA and the AI Act. 
\section{Systemic Risk in the DSA}
\label{sec:dsa}

The DSA introduces a comprehensive framework for platform regulation that centers, inter alia, on the concept of systemic risk, though it approaches this concept differently than AI Act or financial regulation \citep[Sec. 7]{micova2023elements}. While Article 2(c) of the ESRB Regulation provides an explicit definition of systemic risk, just like the AI Act in its Article 3(65) (see below, next section), the DSA embeds the concept throughout its provisions without offering a singular definition; instead it operationalizes the concept through a non-exhaustive list of systemic risks for VLOPs and VLOSEs~\cite{griffin2025politics}.

The DSA employs a quantitative approach based on reach to determine which platforms bear systemic risk obligations. Article 33 establishes that platforms or search engines with average monthly active recipients of 45 million or more in the Union qualify as VLOPs or VLOSEs~\cite{ec2022vlops}.

For those, Article 34(1) DSA establishes the core obligation for risk assessment and mitigation. The provision identifies four specific categories of systemic risk that platforms must assess~\cite{roozenbeek2020susceptibility}. Article 34(1)(a) DSA defines systemic risk in relation to illegal content broadly, as clarified by Recital 80, which extends the concept to illegal activities conducted through platforms and requires a systemic element of dissemination, amplification, or propagation, irrespective of platform terms and conditions.
Next, Article 34(1)(b) obliges VLOPs and VLOSEs to assess actual or foreseeable negative effects on a range of fundamental rights under the Charter, with Recital 81 highlighting risks such as the exploitation of minors, while the DSA controversially presupposes horizontal applicability of these rights between platforms and users.
Third, Article 34(1)(c) targets risks to civic discourse, electoral processes, and public security, thereby recognizing platforms as democratic infrastructure, although the absence of detailed recital guidance reflects persistent Member State disagreement on platform responsibilities in political speech.
Finally, Article 34(1)(d), further specified by Recital 83, addresses systemic risks related to gender-based violence, public health, and minors, and it explicitly acknowledges serious physical and mental harms, such as behavioral addiction and health-related disinformation amplified during crises like the COVID-19 pandemic.
 
Article 34(2) identifies specific factors that platforms must consider in their risk assessments, with Recital 80's emphasis on "rapid and wide" spread providing the analytical framework. The design of recommender systems receives particular attention, as these algorithms determine information flow and can amplify harmful content or create filter bubbles. 

\section{Systemic Risk in the AI Act}
\label{sec:aiact}

Systemic risk in the AI Act forms part of the rules that govern GPAI models under Articles 51 to 56.

\subsection{The Rules in the AI Act}

Article 53 AI Act establishes general transparency and copyright obligations for providers of GPAI models. Only providers of GPAI models with systemic risk (GPAISR) must meet substantial risk management obligations under Art. 55. For this, two sets of rules determine whether a model qualifies as possessing systemic risk. Article 51 and Annex XIII provide the primary classification criteria, while definitions in Article 3 supply interpretative guidance.

Article 51(1) AI Act states that a GPAI model qualifies as having systemic risk if it meets any of the following conditions:

a.	The model has high-impact capabilities evaluated on the basis of appropriate technical tools and methodologies, including indicators and benchmarks.

b.	The Commission determines, either ex officio or following a qualified alert from the scientific panel, that the model has capabilities or an impact equivalent to those in point (a), having regard to the criteria set out in Annex XIII.

While only one of the Annex XIII criteria is connected to compute, a crucial presumption arises that a GPAI model has high-impact capabilities, pursuant to Article 51(1)(a) AI Act, when the cumulative amount of computation used for its training exceeds $10^{25}$ FLOPs (Art. 51(2) AI Act).

This presumption and the wording in Article 51(1) (“any of the following”) at first glance seem to suggest that high-impact capabilities alone suffice for classification under Article 51(1)(a). Article 3(64) AI Act defines high-impact capabilities as capabilities that match or exceed those recorded in the most advanced general-purpose AI models. However, a purposive and systematic analysis of the relevant rules indicates that high-impact capabilities alone cannot determine classification as GPAISR. Without further qualification, any model among the most advanced would qualify as systemically risky even if, for some reason (e.g., near-perfect guardrails), it cannot or does not truly exhibit systemic risk. Rather, Article 51(1) AI Act must be understood as providing that, while “high-impact capabilities” are essential to proving systemic risk, a GPAI model also needs to exhibit “systemic risk” as defined in the AI Act to be designated as GPAISR under Art 51.

Indeed, systemic risk is (confusingly) defined in the AIA quite separately from Article 51. Article 3(65) defines systemic risk as 

“a risk specific to the high-impact capabilities of GPAI models, having a significant impact on the Union market due to their reach or due to actual or reasonably foreseeable negative effects on public health, safety, public security, fundamental rights, or society as a whole, and that can be propagated at scale across the value chain.””.

This definition confirms that high-impact capabilities are necessary but not sufficient. Rather, it contains three elements: The risk must be i)	significant along some dimension (fundamental rights, public health, safety, society as a whole); ii) specific to those high-impact capabilities, and; iii)	prone to propagation at scale across the value chain.

\subsection{The Code of Practice}

The General-Purpose AI Code of Practice (CoP) from July 2025 offers expert-crafted guidance to providers of foundation AI models. Providers who sign and abide by the CoP benefit from a presumption of conformity concerning the covered sections of the AI Act. The Code is a key instrument for understanding systemic risk in the AIA. During CoP negotiations, considerable tension arose as to whether “systemic risk” should effectively concentrate on the more extreme societal or existential risks, which tend to fall into what is known as “AI safety” \cite{bengio2026international}, or extend with equal concern to other groups of risks, e.g. concerning fundamental rights. These might be more likely to manifest here and now, and to affect many vulnerable individuals, but would introduce greater uncertainty and difficulties in calculating costs for providers.

The end result, pro tem, is a compromise. Appendix 1 to the Code acknowledges five sets of distinct but in some cases overlapping risks relevant to systemic risk: risks to public health, to safety, to public security, to fundamental rights, and to society as a whole. These risks are further characterized by essential attributes known from Art. 3(65) (specific to high-impact capabilities, significant market impact, and propagation at scale - see Appendix 1.2.1 ) and contributing characteristics (capability-dependent, reach-dependent, high velocity, compounding effects, difficulty of reversal, and asymmetric impact - see Appendix 1.2.2).
Based on these, a list of specified systemic risks is provided in Appendix 1.4: CBRN; loss of control; cyber offense; and harmful manipulation. These categories are largely, though not exclusively, limited to what might be termed traditional AI safety concerns \cite{bengio2026international}. Notably, it makes almost no reference to fundamental rights except to a very limited extent in harmful manipulation (where notably there is some co-equivalence in material scope with the existing systemic risk provisions of the DSA).

\subsection{Conceptual Critiques}

Our conceptual critiques center on three main aspects: the fundamental rights gap in the Code of Practice; the unnecessary restriction of GPAI models to the ``most advanced models;'' and the equally misguided restriction of systemic risk to instances that have an effect on the Union market.

\subsubsection{The Code and its Fundamental Rights Gap}

Within its framework for identifying systemic risks, the CoP sidelines environmental or fundamental rights risks, such as discrimination and hallucinations~\cite{hacker2024sustainable}. While fundamental rights explicitly appear as a category of potential systemic risks in Appendix 1.1 of the Code and are referenced among capabilities that can trigger such risks in Appendix 1.3.1, they are conspicuously absent from Appendix 1.4's list of "specified systemic risks." The omission of fundamental rights as a specified systemic risk represents a significant gap, particularly given that large-scale discrimination and systematic production of false information (including hallucinations) could pose comparable threats to societal well-being and democratic institutions.

The Code does provide a pathway for fundamental rights concerns to re-enter the risk assessment framework through Measure 2.1(1), which requires providers to identify systemic risks through the structured process that considers all risk types from Appendix 1.1. However, this indirect approach creates uncertainty as to whether risks relating to the environment, discrimination, or hallucinations will in practice be considered by providers and the  AI Office. The focus clearly is on the four specified systemic risks from Appendix 1.4.

\subsubsection{The Act and its Link to ``Most Advanced GPAI Models''}

The AI Act's systemic risk definition contains a conceptual error by tying systemic risk to risks \textit{specific} to high-impact capabilities. The legislative text does not distinguish clearly between systemic risk as a concept and the concrete criteria for determining which models fall within the scope of Article 55 obligations.

Systemic risk, as explained, should refer to significant collective and individual risk, as captured by the first and third element of the Art. 3(65) definition. Yet, this type of risk can arise independently of a model’s level of “advancement.” In our view, smaller or less “advanced” models can also create systemic risks if they produce significant negative effects on public health, safety, security, fundamental rights, or society that can propagate at scale across the value chain. The AI Act’s current language fails to clearly reflect this reality.

There are at least three ways in which the specificity requirement can be read. First, the risk that needs to be specific could be broadly construed as a risk \emph{category}, for example, the support of cyber attacks, disinformation and discrimination campaigns, or bio-terrorism. These risk categories, however, are present as such not only in the most advanced but also in older legacy models \cite{hacker2025generative}. 
Hence, if risk is understood as a risk category, it threatens to exclude many risks types from the concept of systemic risk because they are present not only in the most advanced but also in less capable models.

Second, more narrowly, one could argue that what needs to be specific in the most advanced models is not the risk category, but the very concrete and specific way in which the risk unfolds and manifests itself. Risk, under Art 3(2) AI Act and more generally \cite{gardoni2014scale}, is typically understood as a combination of the probability of the occurrence and severity of harm. If that product is significantly higher for a concrete risk in more advanced than in less advanced models, one may assume specificity. What is specific would not be the absolute risk category, but the more gradual risk \emph{level}. For example, CBRN risks might be present both in legacy and frontier models, but if the latter provide a significantly larger capabilities uplift for potential attackers \cite{krishna2025evaluating,mouton2024operational,kumar2025quantifying}, that would qualify the risk as systemic. There are convincing arguments in favor of such a more nuanced understanding: it aligns with the definition of risk in Article 3 and the description in Recital 110, which mentions that systemic risks increase with model capabilities, suggesting again a gradual perspective. Still, a significant marginal risk increase in the most advanced models is necessary for a risk to qualify as systemic; if the risk is (roughly) the same between the most and less advanced models, it does not count as systemic, even if the risk level itself is high in absolute terms.

A third interpretation would hold that systemic risks need only be specific to high-impact capabilities that are present in – not exclusive to – the most advanced models, such as text production or image generation. Under this reading, a broader spectrum of issues in GPAI models, including hallucinations and large-scale discrimination, could fall under systemic risk. However, the CoP defines model capabilities more narrowly (Appendix 1.3.1, e.g., cyber offense; CBRN capabilities), and the GPAI guidelines issued by the AI Office assume that high-impact capabilities do not include generic generative techniques but are specific or unique to the most advanced models (paras. 34–35 of the Guidelines). Systematically, Art. 51(1) and (2) AI Act tie high-impact capabilities to advanced performance on benchmarks and to compute; this only makes sense if more specific, advanced capabilities are meant, not mundane generative capabilities. The upshot is that, for as much as we would prefer a more open formulation, the second interpretation offered here prevails.

Going forward, systemic risk and models covered by Article 55 should be distinguished in the AI Act; the latter reflect a distinct regulatory choice and may apply to only a subset of the former. One may legitimately decide to limit the obligations under Article 55 to certain types of models that exhibit systemic risk and that meet additional thresholds. Such thresholds could include computational measures such as a minimum number of floating-point operations used during training, a minimum level of capability as measured by benchmarks, a release date after a specified point in time, or the size of the provider company. These thresholds do not arise because other models are incapable of producing systemic risk. Rather, they serve practical policy purposes. These purposes may include sparing small and medium-sized enterprises disproportionate burdens, offering a bright-line rule for regulatory clarity, or focusing enforcement resources on the most likely sources of systemic risk. These legitimate policy considerations for limiting regulatory scope should not be confused with the underlying risk assessment.

The AI Act should state these thresholds explicitly and separate the criteria for systemic risk from the criteria for covered models subject to Article 55 obligations. The current legislative text implies that systemic risk can arise only among the most advanced models. This implication contradicts documented instances in which established and conventional GPAI models have caused severe negative impacts \cite{hacker2025generative,uuk2024taxonomy}. Examples range from sexualized violence through "nudification" ("deep nudes") in non-frontier image models \cite{kopecky2026phenomenon,di2024contemporary,pira2025fake,hawkins2025deepfakes} 
to legacy AI in banking that produces recurring failures through technical debt~\cite{jin2024breaking}.

In fact, the focus on the "most advanced" models creates perverse dynamics. A "moving target problem" emerges as new models cause previously frontier systems to lose their systemic risk classification despite unchanged risks and deployment contexts. During capability plateaus, the definition even becomes entirely fuzzy regarding which models count as "most advanced."

\subsubsection{The Act and the Restriction to the ``Union Market''}

The AI Act's definition of systemic risk contains another problematic limitation: it requires risks to have ``a significant impact on the Union market'' to qualify as systemic. This market-centric framing creates substantial difficulties when assessing risks that primarily affect non-economic interests, which may have profound societal importance -- but not or not primarily for markets. The tension becomes apparent when evaluating discrimination, hallucinations, and environmental harms (see Section~\ref{sec:examples}). 

Large-scale discrimination against protected groups fundamentally violates human dignity and equal treatment principles, yet its "market impact" may be indirect or diffuse. Similarly, AI-generated misinformation that undermines individual personality rights or democratic discourse may not translate readily into market metrics. Environmental degradation from AI's massive energy consumption threatens planetary boundaries but again fits only awkwardly within a market impact framework.

There are two key ways in which the reference to Union market may be interpreted. First, one could narrowly read it as narrow reference to economic or commercial interests. This constraint might stem from the Act's legal basis in Article 114 TFEU, which provides competence for internal market harmonization, and the roots of the AI Act in product safety regulation. Under this reading, the European legislature would have subordinated substantive protection goals to this framing. Yet it would also create a conceptual mismatch between the risks AI poses and the regulatory framework's scope.

In our view, such an interpretation of the market impact requirement represents an outdated regulatory paradigm ill-suited to AI's transformative effects on society, more suited indeed to the financial origins of systemic risk. Systemic risks from AI transcend market boundaries -- they reshape social relations, political processes, and ecological systems. 

Hence, a second and preferable interpretation would broaden the concept of 'Union market.' The definition in Art. 3(65) AI Act seems to imply as much, as it speaks of an impact of the Union market "due to [...] negative effects on [...] fundamental rights". This wording is not conclusive, though, as a separate market impact needs to be shown. This market impact can, however, be read in the context of the AI Act's scope -- which goes vastly beyond economic market transactions, e.g. in Annex III -- and of its aims -- which comprise fundamental rights protection (Art. 1(1) AI Act). In this sense, the Union market is not an unregulated laissez-faire space of economic interaction, but a tightly regulated, constitutionally instituted and legally designed space \cite{}. Under such a perspective, fundamental rights, as core constitutive elements of a legally constructed market, are part of the Union market setting, and have to be considered as part of the systemic risk analysis.   
Hence, a second and preferable interpretation would broaden the concept of 'Union market.' The definition in Art. 3(65) AI Act seems to imply as much, as it speaks of an impact of the Union market "due to [...] negative effects on [...] fundamental rights". This wording is not conclusive, though, as a separate market impact needs to be shown. This market impact can, however, be read in the context of the AI Act's scope -- which goes vastly beyond economic market transactions, e.g. in Annex III -- and of its aims -- which comprise fundamental rights protection (Art. 1(1) AI Act). As scholars from Polanyi onwards have stressed, the market in general and the Union market in particular is not an unregulated laissez-faire space of economic interaction, but a tightly regulated, constitutionally instituted and legally designed space \cite{joerges2014europe,maduro1998we,scharpf1999governing}. Under such a perspective, fundamental rights, as core constitutive elements of a legally constructed market, are part of the Union market setting, and have to be considered as part of the systemic risk analysis.

\section{Comparing Systemic Risk in the DSA and the AI Act}
\label{sec:comparison}

Since systemic risks are mentioned both in the DSA and in the AI Act, the question of their differentiation, but also interaction naturally arises~\cite{hacker2024aiact,helberger2023chatgpt}. This question is all the more pressing because of the increasing emergence of hybrid systems, in which AI models and systems are integrated into platforms and search engines. When search engines like Bing embed generative AI, LinkedIn enhances posts with AI, or X incorporates AI-generated content, the resulting hybrid systems generate risks that neither framework adequately addresses in isolation. This is precisely the model-platform integration risk described in the conceptual part. The recent Grok controversy concerning NCII and child sexual imagery ("deep nudes") is spot on here \cite{bbc_grok_iwf_2026}.

First, we should note that the DSA does not include a restriction of systemic risk to the “most advanced” platforms, unlike the AI Act with its limitation to the “most advanced” GPAI models. While the DSA section only applies to VLOPs and VLOSEs, the moving target problem and the coverage of risks in older platforms, which still maintain VLOP or VLOSE status, does not arise, at least not in the same urgency. 
The DSA, in this sense, is more future-proofed than the nominally much more future-oriented GPAI section of the AI Act.

Second, the AI Act contains, in its systemic risk definition, a market-based logic (significant impact on the Union market) that is entirely foreign to the DSA. This comes as a surprise as the DSA, just like the GPAI rules of the AI Act, is based on Art. 114 TFEU (cf. Recital 3 AI Act). But that did not prompt the DSA framers to restrict systemic risk to Union market effects -- another significant limitation of the AI Act systemic risk category.

Third, the DSA and the AI Act do share an element of reach. Under the AI Act, systemic risk must be capable of propagating down the AI value chain.
By contrast, the DSA threshold concerns a “horizontal” spread of risks among users, as it were, while the AI Act focuses on the “vertical” diffusion into services and applications built with or on top of GPAI models.

Fourth, even further interactions between the DSA and the AI Act systemic risk categories remain. Traditional compliance approaches may treat AI Act and DSA obligations separately -– AI providers assess models for narrowly defined safety while DSA host providers evaluate content moderation and fundamental rights impacts of content on platforms and search engines. Such a siloed approach would fail, however, to capture how platform distribution mechanisms transform AI risks into systemic societal concerns. Hybrid AI-platform systems harbor new or exacerbate systemic risks, as the Grok NCII case industrializing sexual harassment clearly shows. A biased or harmful AI output reaches millions when amplified through platform recommendation algorithms, creating risk magnitudes that likely exceed the sum of individual components. 
 
Recital 118 of the AI Act provides a first insight into that linkage. 
For AI models and systems embedded into VLOPs and VLOSEs, the recital suggests an alignment which collapses AI Act duties onto the DSA: Since VLOPs and VLOSEs are already subject to risk management provisions under the DSA, “the corresponding obligations of [the AI Act] should be presumed to be fulfilled, unless significant systemic risks not covered by [the DSA] emerge and are identified in such models.” Recital 15 of the Commission's Digital Omnibus on AI proposal (COM(2025) 836 final) doubles down on this DSA primacy: ex ante investigations of hybrid GPAI/DSA systems are exclusively relegated to the Commission's DSA team.

Yet, such prioritization of DSA risk mitigation can only work if the mutually reinforcing effects of AI and platform risks are integrated into a combined AIA/DSA risk assessment. Effective governance requires a "reciprocal risk analysis" that examines three interconnected dimensions \cite{hacker2024aiact}. First, platform-specific DSA risks must now account for AI integration effects. Second, AI-specific risks under the AI Act require recontextualization within platform deployment environments. Third, and most critically, emergent risks arise specifically from technological convergence –- new risk categories that neither framework anticipates independently.

For example, Microsoft's assessment of Bing's conversational AI cannot stop at model safety metrics but must consider how search integration affects information discovery patterns and how platform reach transforms model errors into societal phenomena. Conversely, platforms conducting DSA Article 34 assessments must treat AI deployment as a potentially transformative element that modifies all existing risk categories, not merely an additional feature. For example, the integration of AI overviews into traditional search engines might be considered to create a pervasive systemic risk of misinformation (see also Section~\ref{sec:hallucinations}). Only with such interwoven assessments can the premise of Recital 118, that the DSA risk assessment essentially captures all systemic risks from AI, remotely hold.

This has real consequences. The recognition of convergent and amplified risks elevates mitigation obligations beyond traditional approaches. Bias mitigation exemplifies this challenge: while isolated AI systems might address bias through diverse training data and output filters, platform-deployed AI requires additional measures to prevent recommendation algorithms from concentrating biased outputs toward vulnerable populations or amplifying discriminatory patterns through network effects -– even if this comes at the cost of content propagation and “user engagement”. 


\section{Implementing the Frameworks: Examples of Systemic Risk}
\label{sec:examples}

This section examines how the respective rules address systemic risks across three key categories under the DSA, the AI Act and Code of Practice, and our own framework articulated in Section~\ref{sec:conceptual}.



\subsection{Discrimination at Scale}
Discrimination at scale refers to systematic differential treatment of individuals or groups through digital platforms and AI systems based on protected characteristics. The scale element distinguishes such digital discrimination from individual discriminatory acts. A single biased algorithm can impact vast user populations instantaneously, while platform design choices can create structural barriers that systematically exclude or disadvantage protected groups. Examples include facial recognition systems that perform poorly on darker skin tones, content moderation systems that disproportionately flag posts from minority communities, or recommendation algorithms that reinforce gender stereotypes in job advertisements -- or that flood timelines with gendered NCII.

Article 34(1)(b) DSA explicitly identifies discrimination at scale as a systemic risk and specifically mentions "non-discrimination enshrined in Article 21 of the Charter." 

The AI Act, in turn, presents greater challenges. It may address discrimination through its systemic risk framework for general-purpose AI models.

\paragraph{Specificity to most advanced models}
Again, the problem arises if large-scale discrimination is a specific risk of the most advanced models. Research demonstrates complex relationships between model size and bias~\cite{bai2022constitutional,ganguli2023capacity,bender2021dangers,weidinger2022taxonomy,jeong2024bias,an2024measuring,kumar2024investigating,bai2025explicitly,xu2023study}.

The empirical evidence indicates that discrimination at scale constitutes a particular concern for the most advanced GPAI models, though the relationship proves non-linear. Research by Anthropic on Constitutional AI notes that larger models combined with RLHF or RLAIF can reduce certain explicit biases~\cite{bai2022constitutional,ganguli2023capacity}.
However, already the foundational stochastic parrot paper ~\cite{bender2021dangers} suggests that scale often amplifies biases present in training data, the intuition being that more training data also means inviting more of the generally prevalent social and historical biases into the model~\cite{(weidinger2022taxonomy}.
 
The quantitative evidence supports the conclusion that systemic discrimination risks may increase with model advancement. One large study of bias across different models and scales finds that biases are quite model- and context-dependent, and that larger scale does not guarantee more fairness (Jeong et al., 2024). 
In fact, explicit bias still remains an issue in the latest models \cite{an2024large}, and implicit bias actually seems to increase with model size, for example in the Llama and OpenAI GPT families \cite{kumar2024investigating}. 

Overall, the findings concerning the relationship between model size, advancement and bias are mixed~\cite{jeong2024bias,an2024measuring,kumar2024investigating,bai2025explicitly,xu2023study}. Yet, one may additionally argue that the most advanced models are particularly prone to be utilized across a broad area of use cases, enlarging the negative impact of bias vis-à-vis the less advanced models. Overall, while bias exists across model sizes and degrees of advancement, the scale of potential harm makes discrimination a systemic threat specifically linked to the most advanced AI systems. The combination of wider deployment, greater user trust, and persistent subtle biases creates conditions where, in our view, advanced models can perpetuate discrimination at unprecedented scale.

\paragraph{Effect on Union Market}
Next, to qualify as a systemic risk, a significant impact on the Union market must exist, based on a reasonably foreseeable and negative impact on, e.g., fundamental rights (Article 3(65) AI Act). The impact of AI-driven discrimination on the Union market arguably extends significantly beyond fundamental rights violations to create measurable economic distortions. For example, when AI systems introduce bias into employment decisions, they prevent optimal matching between workers and positions. 
In financial services, discriminatory algorithms in credit scoring and insurance pricing restrict capital access for protected groups.


Hence, we conclude that large-scale discrimination does count as systemic risk, even though empirical questions concerning the relationship between degrees of bias and model advancement persist. The Code of Practice seems to support this finding by listing discrimination as one particular risk that must be considered in identifying systemic risks (Appendices 1.1 and 1.3.2 of the Safety and Security Chapter)

\paragraph{Our framework}
In our framework, discrimination rises to systemic risk when it transcends isolated incidents to manifest as large-scale societal harm. Single discriminatory outputs, while problematic, do not constitute systemic risk. However, models trained on skewed data or subject to biased reinforcement learning can perpetuate discrimination across millions of interactions.  
The scale criterion is satisfied when discrimination affects substantial portions of protected groups across the Union. Interconnectedness appears through the reinforcement of stereotypes that influence employment, housing, credit, and social opportunities. Collective harms manifest as marginalized groups face compounded disadvantages that reshape societal structures and political processes. The irreversibility dimension is particularly salient: apologies or model updates cannot undo the corrosive effects of systematic exclusion or disparagement. As discriminatory patterns propagate through the value chain and as downstream applications inherit and potentially amplify biased behaviors, large-scale discrimination, in our view, clearly merits the label of “systemic risk”.

\subsection{Information Pollution Through Hallucinations}
\label{sec:hallucinations}

``Hallucinations'' in AI systems refer to outputs that contain factually false or misleading information~\cite{binns2025reputation,magesh2025hallucination}. Hallucinations are inevitable in current LLMs~\cite{xu2024hallucination}. Studies show varying hallucination rates~\cite{chelli2024hallucination,zhao2024wildhallucinations,wei2024longform,nielsen2025hallucinations}. 

\paragraph{DSA}
The DSA captures AI hallucinations through two systemic risk categories when they manifest on digital platforms. First, hallucinations that generate misinformation about electoral candidates, voting procedures, or political events fall under Article 34(1)(c)'s protection of democratic processes. Second, creating false statements about individuals constitutes a risk to personality rights, which fall under the fundamental rights category in Article 34(1)(b).

Note, however, that these risks are covered by the DSA only when VLOPs and VLOSEs integrate generative AI capabilities into their services, as platforms themselves are not necessarily prone to hallucinations (and AI models not integrated into platforms are not per se regulated by the DSA). However, such hybrid systems proliferate rapidly across the digital ecosystem. Search engines now embed generative AI features, as demonstrated by ChatGPT Search and Perplexity, while social media platforms like Twitter/X and LinkedIn integrate LLMs for automated content generation. This convergence of traditional platforms with AI capabilities expands the DSA's relevance for addressing hallucination rates and pertinent risks.
The European Commission is rightly considering designating ChatGPT Search as a VLOP or VLOSE, for example \cite{lorente2026between}.

\paragraph{AI Act}
The AI Act's Code of Practice explicitly recognizes hallucinations as a potential systemic risk that requires assessment and mitigation. Appendix 1.3.2 of the Safety and Security Chapter identifies hallucinations among the key risks and model propensities that providers of general-purpose AI models must evaluate.
 
Hallucinations persist even in advanced models \cite{zhao2024wildhallucinations,chelli2024hallucination,magesh2025hallucination}. 
Studies show hallucination rates of 40\% for GPT 3.5 and 29\% for GPT 4.0 (Chelli et al., 2024), or 17\% and 33\% for specialized legal domain models trained by LexisNexis (Lexis+ AI) and Thomson Reuters (Westlaw AI-Assisted Research and Ask Practical Law AI), respectively (Magesh et al., 2025). 
However, research also demonstrates that hallucination rates decrease with model size and successive generations \cite{wei2024longform,nielsen2025hallucinations}, even though it persists at lower levels. 
Larger models with more parameters generally exhibit better factual accuracy due to increased capacity to encode knowledge from training data. 

Here, the paradox emerges clearly: while hallucination rates decrease with model advancement, they remain prevalent enough in the most sophisticated systems to pose significant risks. Yet, it is difficult to square the  AI Act's GPAI rules with this risk. On the one hand, one might argue that the systemic risk concept does not explicitly capture hallucinations as a distinct category requiring specific regulatory measures as hallucination rates are not higher in, and hence not specific to, the most advanced models – quite the inverse. On the other hand, one might argue that despite lower hallucination rates, the overall risk of harm flowing from hallucinations may be significantly higher in the most advanced models. For example, hallucinated information may be rendered more convincingly -- linguistically and stylistically -- and also with seemingly correctly but actually misleading sources. Hence, recipients may be more likely to act on it in harmful ways. But this depends on the concrete scenario and the ways in which information is presented, and leaves a narrow road to the recognition of hallucination under the systemic risk framework, at best.

If this road cannot be taken, a regulatory gap arises that leaves a documented fundamental risk without targeted legal obligations, despite its recognition in the Code of Practice and its potential for widespread harm through misinformation propagation.

At least, hallucinations can and must be considered under the risk management framework of Article 9 AI Act, in high-risk AI systems. However, many typical applications of LLMs in chatbots do not fall within any of the high-risk activities listed in the AI Act (Annexes I and III). Overall, hallucinations are, therefore, one prime example where the systemic risk categorization between the DSA and the AI Act may markedly diverge.

\paragraph{Our framework}
Under our framework, hallucinations can and generally do count as systemic risk. The severity of AI hallucinations –- confident generation of false information –- varies with context and scale. Isolated errors, such as incorrect birthdays or minor factual mistakes, remain localized problems. Even serious individual harms, like the Norwegian man erroneously connected to child murder by ChatGPT, may not constitute systemic risks if they remain exceptional.

However, hallucinations become systemic when they poison society's information ecosystem at scale \cite{wachter2024legal}.
Widespread generation of plausible-sounding falsehoods meets the scale criterion through mass exposure across the Union. Interconnectedness emerges as false information influences decisions, shapes public opinion, and enables coordinated disinformation campaigns. The complexity of language models makes comprehensive accuracy impossible to guarantee based on the current transformer paradigm. Collective harms arise when societal trust in information sources erodes (including evidence used in courts), democratic discourse degrades, and shared factual foundations dissolve. While individual falsehoods might be corrected, the cumulative effect on information integrity proves difficult to reverse. These risks cascade through the value chain as applications built on hallucination-prone models spread misinformation across diverse contexts. Overall, hallucinations are a clear candidate for systemic risk, but the AI Act struggles to capture them via its GPAI rules.

\subsection{Climate and Environmental Impacts}

AI has significant potential for reducing emissions~\cite{rolnick2022tackling,taddeo2021artificial}, but the current tendency is for AI to generate more demand for energy~\cite{iea2025energy,kaack2022aligning,luccioni2024power,luccioni2025efficiency}. The rapid growth has led to significant climate and environmental impacts~\cite{oecd2022measuring,gelenbe2022measurement,nartey2025environmental,hsu2022water,ebert2024ai,hacker2024sustainable,ruschemeier2026climate}.

\paragraph{DSA}

Emerging scholarship argues that climate-related harms threaten the societal interests the DSA aims to protect~\cite{griffin2023climate,kaesling2025sustainability}.
Recent legal and policy analyses indicate that environmental harm – including contributions to climate change – can fall within the scope of “systemic risks” under Article 34 of the Digital Services Act~\cite{griffin2023climate,kaesling2025sustainability}. Although the DSA does not explicitly cite environmental impacts as a standalone risk category, its requirements for VLOPs and VLOSEs to assess and mitigate systemic risks are phrased broadly enough to capture negative effects on public health, physical and mental well-being, and fundamental rights, all of which can be significantly affected by environmental degradation and climate change. Indeed, a habitable planet is precisely a prerequisite for the enjoyment of any of these rights and freedoms \cite{hacker2024sustainable}. Moreover, an increasing number of cases reconfigures environmental interests directly as fundamental rights issues, for example through the lens of the right to live and the right to health~\cite{gera2024environmental,hartmann2022protecting,vanzeben2021role}.

Hence, climate-related and other environmental harms, both direct (e.g., energy and water use of digital platforms) and indirect (e.g., facilitating environmentally damaging behaviors), threaten the societal interests the DSA aims to protect~\cite{griffin2023climate,kaesling2025sustainability}. Accordingly, providers are expected to take reasonable measures to minimize their environmental footprint -- such as enhancing energy efficiency or reducing resource usage -- as part of their Article 34 risk mitigation obligations.

\paragraph{AI Act}

How about the AI Act? In our view, environmental damage and climate change contributions qualify as systemic risks under the AI Act due to the disproportionate resource consumption of advanced AI models \citep[][see also]{hacker2024sustainable}. The computational requirements typically scale with model size~\cite{luccioni2024power,patterson2021carbon,sanchezm2025green,strubell2019energy,li2023making}. 
Water consumption follows similar patterns \cite{li2023making}. 
This is also recognized in the Code of Conduct, where environmental consequences are listed as one type of risk to consider (Appendix 1.1 of the Safety and Security Chapter). While environmental harm and climate effects are also caused by less advanced models, they are disproportionately greater in the most advanced models, particularly also the so-called "reasoning" models that spend significantly more time, and hence energy, on inference compute \cite{ai_energy_score_v2,ebert2026global}.

The effect on the Union market manifests through multiple pathways documented by the European Environment Agency pathways~\cite{eea2024climate} and the European Central Bank~\cite{alogoskoufis2021climate}. 
This convergence of environmental impacts with market effects, in our view, establishes clear grounds for treating climate contributions as systemic risks requiring regulatory intervention under the AI Act.

\paragraph{Our framework}
Environmental harm qualifies as systemic risk under our framework. The scale criterion is met through AI's contributions to climate change across all Member States via massive energy and water consumption. Interconnectedness is particularly pronounced: climate impacts affect energy availability for AI systems, while a hotter planet increases cooling demand, which in turn drives up energy consumption. Collective harms manifest as ecosystem degradation and public health impacts. Environmental damage epitomizes irreversibility: atmospheric carbon persists for centuries, and climate tipping points designate permanent changes.

\section{Overarching Lessons and Policy Proposals}
\label{sec:lessons}

The emergence of systemic risk as a regulatory concept across financial regulation, the Digital Services Act, and the AI Act reveals both common principles and domain-specific adaptations.

\paragraph{Comparative Risk Characteristics}

AI systemic risks combine elements of both financial and platform risks. Like financial risks, they can cascade through technical systems and create sudden disruptions. As platform risks, they shape information environments and human behavior in ways that persist beyond immediate incidents. Like both financial and platform risk~\cite{micova2023elements}, they can be triggered by external or internal sources. 

\paragraph{The Definition of Systemic Risk: beyond the Most Advanced Models}
The three frameworks define systemic risk very differently, though: financial regulation uses precise technical definitions; the DSA avoids definition altogether in favor of a pragmatic, non-exhaustive list of risk categories; the AI Act offers a formal but problematic definition tied to "high-impact capabilities" and "the most advanced" models. In our view, the AI Act's over-specification creates conceptual confusion and undermines regulatory effectiveness. The speed of AI development compounds this problem, as risk profiles evolve rapidly from models to systems and agents; a definition tied to the current model frontier may become outdated before enforcement mechanisms can adapt \cite{hacker2026pragmatic}. Hence, the restriction to the most advanced models should be abandoned, as the analysis of hallucinations shows.

\paragraph{Beyond the Market Paradigm} Future revisions of the AI Act should also move more explicitly beyond the constraint that systemic risks must have a "significant impact on the Union market." Fundamental rights, democratic values, and environmental sustainability ultimately cannot, and should not, be reduced to market effects \cite{weatherill2011limits,kumm2006constitutionalising,hervey2001community}. 
More specifically, based on the current AI Act, guidelines or a new Recital should specify that a 'Union market impact' does not denote a purely economic effect, but also comprises effects on the constitutional and legal boundaries of the market, including fundamental rights. Going forward, the Union market impact should be dropped from Article 3(65); 'actual or reasonably foreseeable negative effects on public health, safety, public security, fundamental rights, or the society as a whole' are enough of a filter to trigger, jointly with the other elements, the systemic risk provisions of the AI Act.

\section{Conclusion}

This paper makes several novel contributions to the systemic risk literature. The concept of systemic risk has a long pedigree in financial regulation, where its defining feature lies in the interconnectedness of actors and the cascading failures that follow. This paper has sought to update that core insight for the governance of AI and digital platforms. We propose a conception of systemic risk that captures the specific propagation mechanisms of these technologies and that extends beyond the narrow thresholds used in recent legislation.

We also advance several policy proposals. Regulators should broaden the definition of systemic risk in the AI Act to encompass risks beyond frontier capabilities. Risk assessments under both the AI Act and the DSA should explicitly address collective harms such as discrimination, systematic misinformation, and climate effects, particularly in hybrid AI-platform systems. Finally, systemic risk obligations should be designed with flexibility to adapt to new forms of technological convergence without creating regulatory gaps, particularly without an unnecessary restriction to phenomena with EU market effects. Overall, integrating insights from finance, algorithmic fairness, and climate science, and foregrounding novel AI-specific pathways to systemic disruption, we hope to offer a framework for AI and platform governance that is both conceptually rigorous and practically oriented.

\newpage
\bibliographystyle{ACM-Reference-Format}
\bibliography{main}

\appendix

\section{Selected Definitions of Systemic Risk}

\begin{table}[htbp]
    \centering
    \caption{Selected definitions of systemic risk}
    \label{tab:def}
    \begin{tabularx}{\textwidth}{lX}
        \toprule
        \textbf{Source} & \textbf{Definition} \\
        \midrule
        \citet{renn2022systemic} & ``Systemic risks are characterized by high complexity, multiple uncertainties, major ambiguities, and transgressive effects on other systems outside of the system of origin.'' \\
        \addlinespace
        \citet{helbing2013globally} & ``Systemic risk is the risk of having not just statistically independent failures, but interdependent, so-called `cascading' failures in a network of N interconnected system components.'' \\
        \addlinespace
        \citet{schwarcz2008systemic} & ``The probability that cumulative losses will occur from an event that ignites a series of successive losses along a chain of [financial] institutions or markets comprising... a system.'' \\
        \addlinespace
        \citet{kaufman2003what} & ``The risk or probability of breakdowns in an entire system, as opposed to breakdowns in individual parts or components.'' \\
        \addlinespace
        \citet{huimin2021understanding} & ``Systemic risk induced by climate change is a holistic risk generated by the interconnection, interaction, and dynamic evolution of different types of single risks, and its fundamental, defining feature is cascading effects. The extent of risk propagation and its duration depend on the characteristics of the various discrete risks that are connected to make up the systemic risk.'' \\
    \addlinespace
  \citet{bloomfield2012systemic} & We consider a serious nuclear incident that has the potential for the release of radioactivity with associated plant damage as a “systemic event” and hence make the link to a financial market crash: an event that both damages the market and also potentially impacts the wider financial system and the broader economy. \\
      \addlinespace
  \citet{davis1992debt} & 
 The most serious direct consequences of systemic risk involve "disrupt[ing] the payments mechanism and capacity of the system to allocate capital." \\
       \addlinespace
        \bottomrule
    \end{tabularx}
\end{table}

\section*{Generative AI Usage Statement}

We have not used generative AI for writing this paper.

\end{document}